\documentclass[]{emulateapj-rtx4}

\usepackage{natbib,color} 
\bibpunct{(}{)}{;}{a}{}{,}

\DeclareRobustCommand{\ion}[2]{%
\relax\ifmmode
\ifx\testbx\f@series
{\mathbf{#1\,\mathsc{#2}}}\else
{\mathrm{#1\,\mathsc{#2}}}\fi
\else\textup{#1\,{\mdseries\textsc{#2}}}%
\fi}

\shorttitle{Center-to-Limb Variation of the Inverse Evershed Flow}
\shortauthors{Beck, C.; Choudhary, D.P.; Ranganathan, M.}

\begin{document}
\title{Center-to-Limb Variation of the Inverse Evershed Flow}


\author{C. Beck}
\affil{National Solar Observatory (NSO), Boulder, USA}
\author{D.P. Choudhary}
\affil{Department of Physics \& Astronomy, California State University, Northridge, USA}
\author{M. Ranganathan}
\affil{Department of Physics \& Astronomy, California State University, Northridge, USA}




\begin{abstract}
We present the properties of the inverse Evershed flow (IEF) based on the center-to-limb variation of the plasma speed and loop geometry of chromospheric superpenumbral fibrils in eleven sunspots that were located at a wide range of heliocentric angles from 12$^\circ$ to 79$^\circ$. The observations were acquired at the Dunn Solar Telescope in the spectral lines of H$\alpha$ at 656\,nm, \ion{Ca}{ii} IR at 854\,nm and \ion{He}{i} at 1083\,nm. All sunspots display opposite line-of-sight (LOS) velocities on the limb and center side with a distinct shock signature near the outer penumbral edge. We developed a simplified flexible sunspot model assuming axisymmetry and prescribing the radial flow speed profile at a known loop geometry to replicate the observed two-dimensional IEF patterns under different viewing angles. The simulated flow maps match the observations for chromospheric loops with 10–-20\,Mm length starting at 0.8--1.1 sunspot radii, an apex height of 2--3\,Mm and a true constant flow speed of 2-–9 km\,s$^{-1}$. We find on average a good agreement of the simulated velocities and the observations on elliptical annuli around the sunspot. Individual IEF channels show a significant range of variation in their properties and reach maximal LOS speeds of up to 12\,km\,s$^{-1}$. Upwards or downwards directed flows do not show a change of sign in the LOS velocities for heliocentric angles above 30$^\circ$. Our results are consistent with the IEF being caused by a siphon flow mechanism driving a flow at a constant sonic speed along elevated loops with a flattened top in the chromosphere.
\end{abstract}

\section{Introduction}
Sunspots are large coherent magnetic structures embedded in the solar interior and atmosphere that extend from below the photosphere to coronal heights and display various structures and different atmospheric features \citep{solanki2003}. Sunspots have three distinct boundaries of increasing radius in the photosphere: the umbra, the penumbra and the superpenumbra or moat. The dark umbra usually seamlessly connects to the penumbra that exhibits thin elongated  filaments in the photosphere, while the connection  to the superpenumbral boundary is visible through similar elongated chromospheric fibrils. There are no obvious structures connecting the penumbra and the superpenumbra in the photosphere in most cases, except for the bright points and moving magnetic features (MMFs) that move radially away from the sunspot in the moat region \citep{sheeley1969,harvey+harvey1973,ravindra+etal2004,beck+etal2007}. Some of the MMFs originate deep in the penumbra and remain connected to it \citep{cabrerasolana+etal2006,ma+etal2015}. The superpenumbral boundary demarcated by the end of the moat flow and the systematic radial motion of the MMFs depends on the size and age of the sunspot \citep{sobotka+roudier2007,verma+etal2018}.

The narrow filaments in the photosphere and the fibrils in the chromosphere are assumed to trace the magnetic field lines connecting the central region of the sunspot with its surroundings \citep{frazier1972,zirin1972,schad+etal2013,beck+choudhary2019}. Plasma flows resulting in an exchange of mass and energy along such magnetic field lines \citep{bellot+etal2003,bethge+etal2012} might be consequential for the stability of these structures, especially higher up in the solar atmosphere \citep{jenkins+etal2018,williams+taroyan2018,li+peter2019}. For sunspots, the mass motions in the photosphere and chromosphere happen mostly in the form of the Evershed and inverse Evershed flow \citep{evershed1909,evershed1910} along horizontal and inclined field lines. In the current series of papers, we investigate the properties of the chromospheric inverse Evershed flow (IEF) that carries mass and energy towards the sunspot. 

The IEF is channeled by the chromospheric fibrils that mostly connect the penumbral and superpenumbral boundaries with only few of them extending to remote magnetic elements \citep{beck+etal2014,ma+etal2015}. It has been predicted and observed that the IEF reaches super-sonic speeds near the sunspot creating shock fronts \citep{thomas+montesinos1991,georgakilas+etal2003,choudhary+beck2018}. However, the speed of the flows and the amount of material transported by them depends on the magnetic field configuration that can be studied by determining the geometrical properties of the IEF channels. 

In this paper, we employ the center-to-limb variation (CLV) of the IEF to study the structure of these chromospheric flow channels. The CLV technique has been successfully employed to investigate a number of properties of sunspots and solar active regions. For instance, the CLV of the broadband circular and linear polarization was used to determine the geometrical orientation of the magnetic field in sunspots by \citet{illing+etal1974}.  \citet{mattig1969} used the CLV in brightness to derive the structure of the vertical temperature gradient in sunspots and \citet{wilson1969} to determine the umbral elevation, while \citet{alissandrakis+kundu1984} were able to determine coronal temperatures and magnetic fields in a sunspot from a CLV study of radio emission. To better understand the chromospheric flow mechanism of the IEF, geometric parameters such as the fibril length, apex height and flow speed of the fibrils harboring the flows are important.  A study of the CLV of the IEF offers an opportunity to determine those parameters by providing different viewing angles on the topology of the flow field.

Section \ref{secobs} describes the observations used. Our analysis methods are explained in Section \ref{secana}. The analysis results are given in Section \ref{secres}. Section \ref{secdisc} discusses the findings, while Section \ref{secconcl} provides our conclusions.

\section{Observations}\label{secobs}
\begin{figure*}
$ $\\$ $\\
\centerline{\resizebox{12cm}{!}{\includegraphics{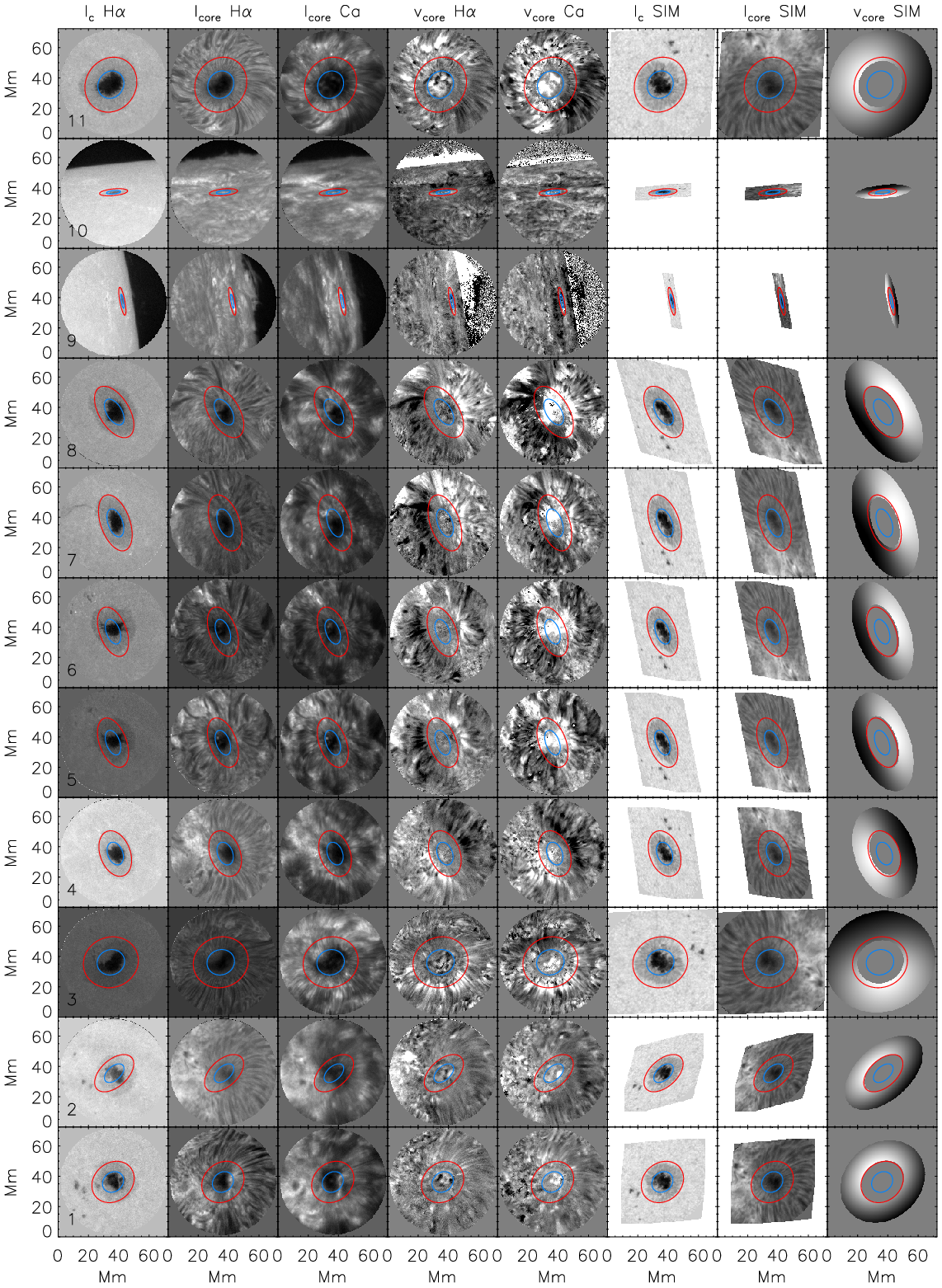}}\hspace*{0.5cm}\includegraphics{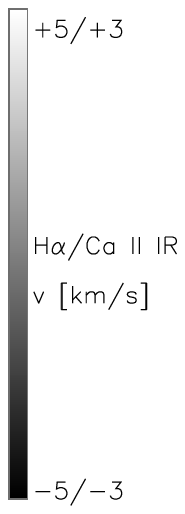}}$ $\\$ $\\
\caption{Overview images of the observations. Left to right: observed continuum intensity I$_c$, H$\alpha$ and \ion{Ca}{ii} IR line-core intensity, H$\alpha$ and \ion{Ca}{ii} IR line-core velocity, synthetic continuum intensity, H$\alpha$ line-core intensity, and velocity from the IEF simulator. Bottom to top: observations Nos. 1--11. The blue (red) ellipses indicate the outer umbral (penumbral) boundary in the IEF simulator. The IEF simulator was run with a fixed apex height of 1\,Mm and the best-fit values for the inner end point and the flow velocity as described in Section \ref{sim_results}.}\label{fig1}
\end{figure*}

We acquired 11 data sets of seven different sunspots at a variety of heliocentric angles with the Interferometric BIdimensional Spectrometer \citep[IBIS;][]{cavallini2006,reardon+cavallini2008} at the Dunn Solar Telescope \citep[DST;][]{dunn1969, dunn+smartt1991} in its spectroscopic mode. All IBIS observations covered the chromospheric spectral lines of H$\alpha$ at 656\,nm and \ion{Ca}{ii} IR at 854.2\,nm with a non-equidistant spectral sampling of about 30 wavelength points \citep[see][their Figure 3]{beck+choudhary2020}. The field of view (FOV) of IBIS was circular with about 95$^{\prime\prime}$ diameter at a spatial sampling of about 0\farcs1 per pixel. We selected a single spectral scan with the best seeing conditions from each IBIS time series. The first five columns of Figure \ref{fig1} show overview images of the continuum and line-core intensity and the line-core velocity of the IBIS observations.

For most observations, additional spectropolarimetric data are available in the same or other spectral lines such as \ion{He}{i} at 1083\,nm that were taken with the SPectropolarimeter for Infrared and Optical Regions \citep[SPINOR;][]{socasnavarro+etal2006} or the Facility InfraRed Spectropolarimeter \citep[FIRS;][]{jaeggli+etal2010}. Table \ref{obstable} lists the dates of the observations, the wavelengths observed by each instrument and the heliocentric angle of each data set. More details on the settings such as exposure or integration times can be found at \href{https://www.nso.edu/telescope/dunn-solar-telescope/cbeck/default.html}{https://www.nso.edu/telescope/dunn-solar-telescope/cbeck/default.html}. The ground-based data are complemented by co-aligned line-of-sight (LOS) magnetograms from the Helioseismic and Magnetic Imager \citep[HMI;][]{scherrer+etal2012} and 1700\,{\AA} images from the Atmospheric Imaging Assembly \citep[AIA;][]{lemen+etal2012} on-board the Solar Dynamics Observatory \citep[SDO;][]{pesnell+etal2012}. 
\begin{table}
\caption{List of Observations$^1$}\label{obstable}
\begin{tabular}{c|ccccc}
No. & Date & IBIS & SPINOR & $\theta$ &DST $(x,y)$ \cr
    &  & $\lambda$ [nm] &  $\lambda$ [nm] & deg & arcsec \cr\hline\hline
1 & 2014/03/13 & 656,854 & 656,854,1083 & 12 & -195,68\cr
2 & 2014/03/14 & 656,854 & 656,854,1083 & 19 & -233,222\cr\hline
3 & 2014/03/17 & 656,854 & 656,854,1083 & 22 & 348,-93\cr
4 & 2014/03/21 & 656,854 & 656,854,1083 &  45 & 133,673\cr\hline
5 & 2014/11/15 & 656,854 & 656,854,1083  &  53 & -47,-766\cr
6 & 2014/11/15 & 656,854 & 656,854,1083  &  53 & -62,-757\cr\hline
7 & 2015/09/15 & 656,854 & FIRS: 1083 &55  & -138,-767\cr
8 & 2015/09/16 & 656,854 & FIRS: 1083 & 43 & -199,-614\cr\hline
9 & 2016/06/15 & 656,854 & -- & 79 & -69,926\cr\hline
10 & 2016/06/17 & 656,854 & -- & 76 & 862,308\cr\hline
11 & 2016/06/17 & 656,854 & FIRS: 1083 &  19 & -227,-197\cr
\end{tabular}
1: Horizontal lines separate observations of the same sunspot at different times or dates.\\
\end{table}

\section{Data Analysis}\label{secana}
\subsection{Bisector Analysis of Chromospheric Spectra}
We applied a bisector analysis to the IBIS spectra in H$\alpha$ at 656\,nm and \ion{Ca}{ii} IR at 854\,nm to retrieve the LOS velocity at different line depths \citep[see, e.g.,][and references therein]{beck+choudhary2020,gonzalezmanrique+etal2020}. We used the bisector velocity at 93\,\% line depth as a measure of the chromospheric velocity in both spectral lines, while the intensity value at the central wavelength that was sampled in the observations is labeled the line-core intensity. 

The zero point of the velocity scale was defined separately for each observation as the average velocity value across the full IBIS aperture. This velocity calibration or forcing the average umbral velocity to zero \citep{cabrerasolana+etal2007,loehner+etal2018} gave nearly identical results within $\pm$0.5\,km\,s$^{-1}$. For the two observations Nos.~9 and 10 at large heliocentric angles we used the latter approach because of the spurious velocities off the solar limb. 

\begin{figure}
\resizebox{8.8cm}{!}{\includegraphics{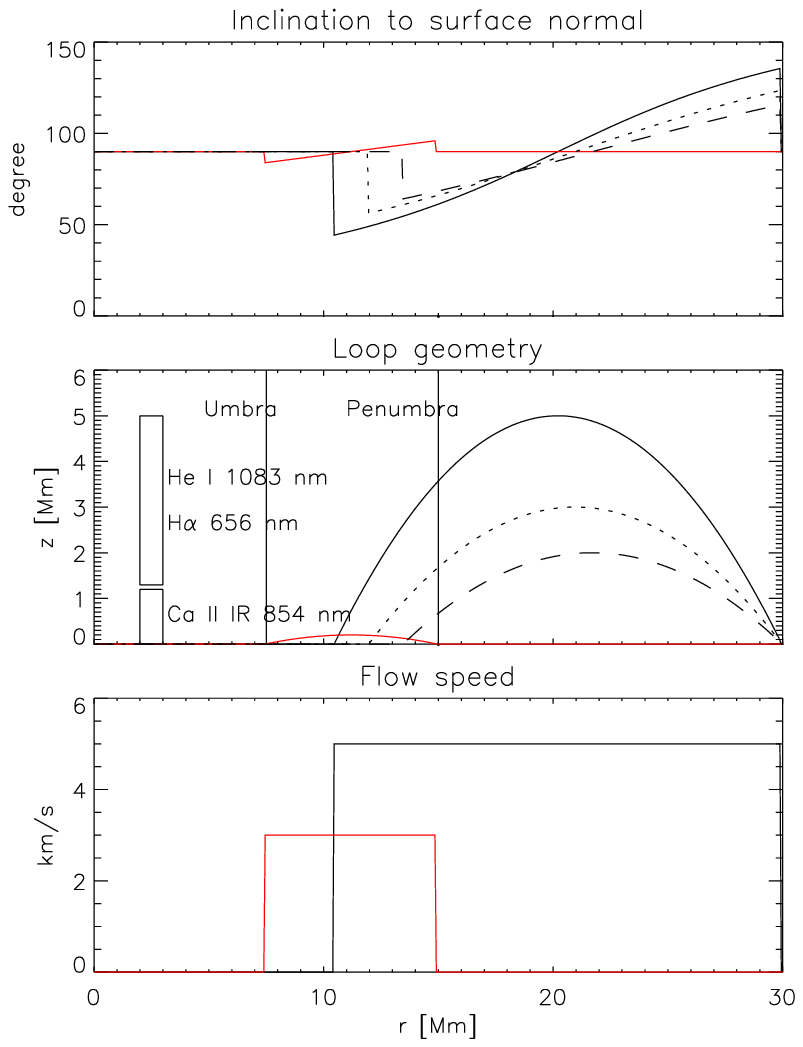}}$ $\\$ $\\
\caption{Settings and geometry in the EF/IEF simulator. Bottom to top: radial flow speed profile, loop geometry and inclination to the local vertical. Red/black lines: EF/IEF channels. The dashed, dotted and solid black lines in the upper two panels indicate IEF channels with 2, 3 and 5\,Mm apex height starting at $r_0 = 0.7, 0.8$ and 0.9. The approximate formation heights of the whole \ion{Ca}{ii} IR spectral line and the line-cores of H$\alpha$ and \ion{He}{i} at 1083\,nm are indicated by rectangles at the left-hand side of the middle panel.}\label{fig2}
\end{figure}

\begin{figure}
$ $\\$ $\\
\centerline{\hspace*{1.cm}\resizebox{7.8cm}{!}{\includegraphics{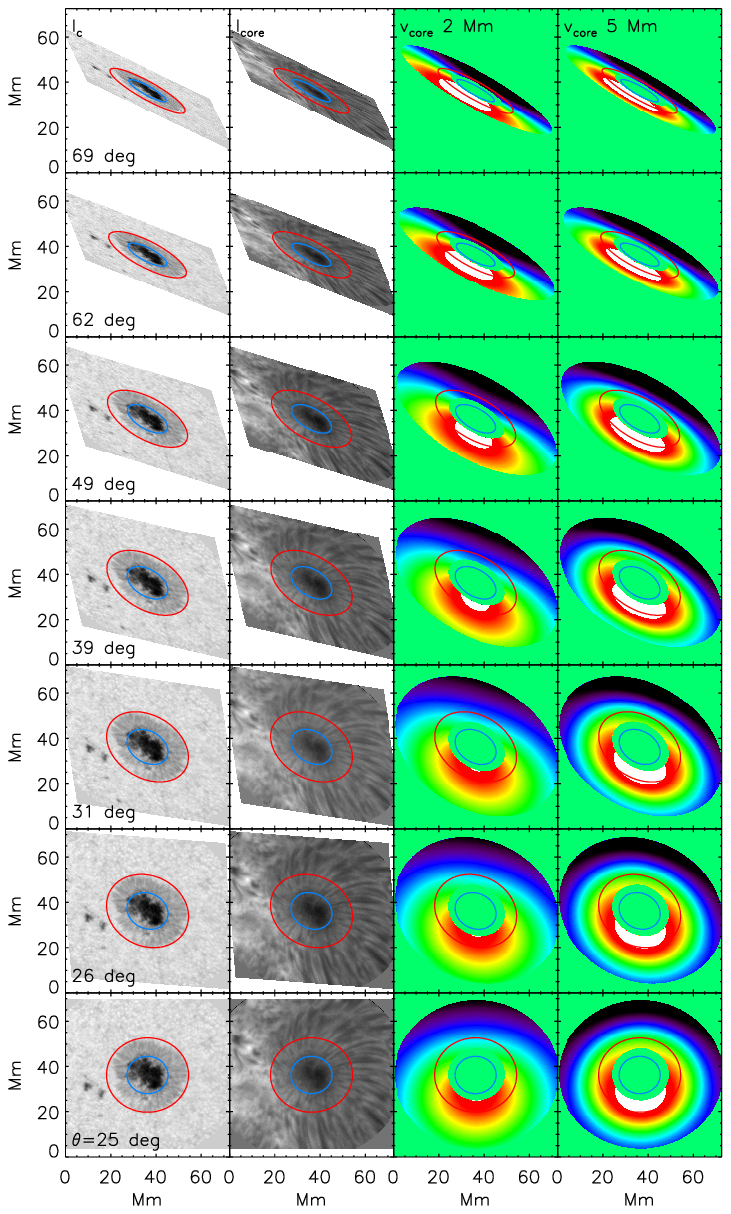}}}$ $\\$ $\\$ $\\$ $\\
\centerline{\hspace*{1cm}\resizebox{4.4cm}{!}{\includegraphics{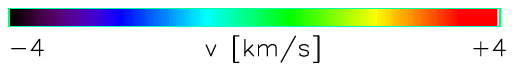}}}
\caption{Example of a CLV run with the IEF simulator. Left to right: projected synthetic continuum intensity, H$\alpha$ line-core intensity, LOS velocities for an apex height of 2\,Mm and 5\,Mm. The heliocentric angle increases from bottom to top for a sunspot located in the north-west solar quadrant. The blue (red) ellipses indicate the projected umbral (penumbral) boundary.}\label{fig3}
\end{figure}
\subsection{Derivation of Umbral Field Strength}
We estimated the umbral field strength $B$ using the wavelength separation between the Stokes V components of the \ion{Si}{i} line at 1082.7\,nm wherever suited data was available (see Table \ref{obstable}). The effective Land{\'e} coefficient of this line is 1.5, which is sufficient to be outside the weak-field limit for $B> 2000$\,G \citep[][their Figure 7]{beck+etal2007}. The error in the determination is assumed to be less than 100\,G \citep{rezaei+etal2012,rezaei+etal2015}. We masked the umbral area by a threshold in the continuum intensity and calculated the average ($<B>$) and maximal field strength ($B_{max}$) in the umbra for each sunspot. 

\subsection{Evershed and Inverse Evershed Flow Simulator}
For comparison to the observed LOS velocities, we constructed a flexible geometrical model of EF and IEF flows based on arched loops in an axisymmetric sunspot with an umbral radius of 7.5\,Mm, a penumbral radius of 15\,Mm and a moat cell of 30\,Mm with a radial sampling of 60\,km. The corresponding ratios of the three characteristic sunspot regions in terms of the umbral radius thus are 1:2:4. The magnetic field lines that harbor the EF are modeled to extend from the outer umbral to the outer penumbral boundary with a parabolic shape with the apex height $AP_{EF}$ of the EF loops as free parameter (see middle panel of Figure \ref{fig2}). With the EF loop length $L_{EF}$ of 7.5\,Mm, the loop height $h_{EF}(r)$ as a function of radial distance is then given by \citep[see also][his Section 3.2]{maltby1975}:
\begin{eqnarray}
h_{EF}(r_{center} + r) = AP_{EF} -  \frac{4 AP_{EF}}{L_{EF}^2} \cdot r^2
\end{eqnarray} 
with $r = -L_{EF}/2$ to $+L_{EF}/2$ and $r_{center} = 11.25$\,Mm. 

The inclination to the local surface normal $\gamma_{EF}(r)$ is then given through
\begin{eqnarray}
\frac{dh}{dr} = - \frac{8 AP_{EF}}{L_{EF}^2}\cdot r \,\,\,\mbox{by}\\ 
\gamma_{EF}(r) = \arctan (\frac{dh}{dr})\,.
\end{eqnarray}

For the loops relevant for the IEF channels, the outer end is set to be at the end of the moat cell, while the location of the inner end $r_0$ can be freely selected inside or outside the sunspot in fractions of the outer penumbral radius. Together with the apex height of the IEF channel $AP_{IEF}$, a variety of loop geometries can be realized that differ in the inclination of the loop $\gamma_{IEF}$ (top panel of Figure \ref{fig2}) and its exact location relative to the (pen)umbral boundary. The only difference to the EF is that now also $r_0$ is a free parameter that changes the length of the loops $L_{IEF} = (30-r_0 \times 15)$\,Mm and the central position $r_{center}$. 

The flow speed of the (I)EF channels $v_{(I)EF}(r)$ is a free parameter that was chosen to be constant with radial distance for simplicity (bottom panel of Figure \ref{fig2}), but could have any specific desired radial shape to mimic acceleration, deceleration, or both.  

The corresponding LOS velocity $v^{LOS}(x,y)$ all across a simulated 1k x 1k sunspot image  is then calculated following \citet[][their Equation (1)]{schlichenmaier+schmidt2000} by
\begin{eqnarray}
v^{LOS}_{(I)EF}(x,y) = v_{(I)EF}(r) \cdot [ \sin \theta\, \sin \Phi\, \sin \gamma_{(I)EF}(r)\nonumber\\ + \cos \Phi \, \cos \gamma_{(I)EF}(r)]\,,
\end{eqnarray}
with $[x^\prime,y^\prime] = [x-500,y-500]$ and $r = \sqrt{ (x^\prime)^2 + (y^\prime)^2}$, while $\theta$ is the heliocentric angle, $\Phi = \arctan(y^\prime/x^\prime)$ is the azimuth angle around the sunspot center, and $v_{(I)EF}$ and $\gamma_{(I)EF}(r)$ are as defined above. 

\begin{figure*}
\begin{minipage}{5cm}
\centerline{\resizebox{5cm}{!}{\includegraphics{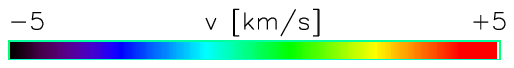}}}$ $\\
\resizebox{5cm}{!}{\includegraphics{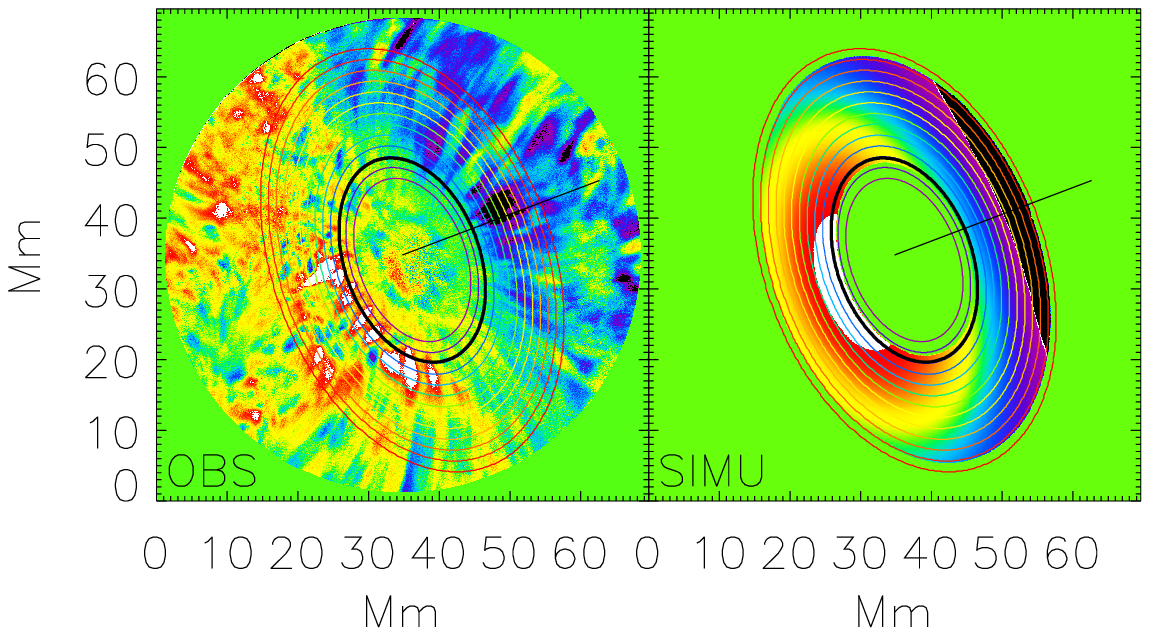}}\\$ $\\$ $\\
\end{minipage}\hspace*{.6cm}
\begin{minipage}{12.cm}
\resizebox{12cm}{!}{\includegraphics{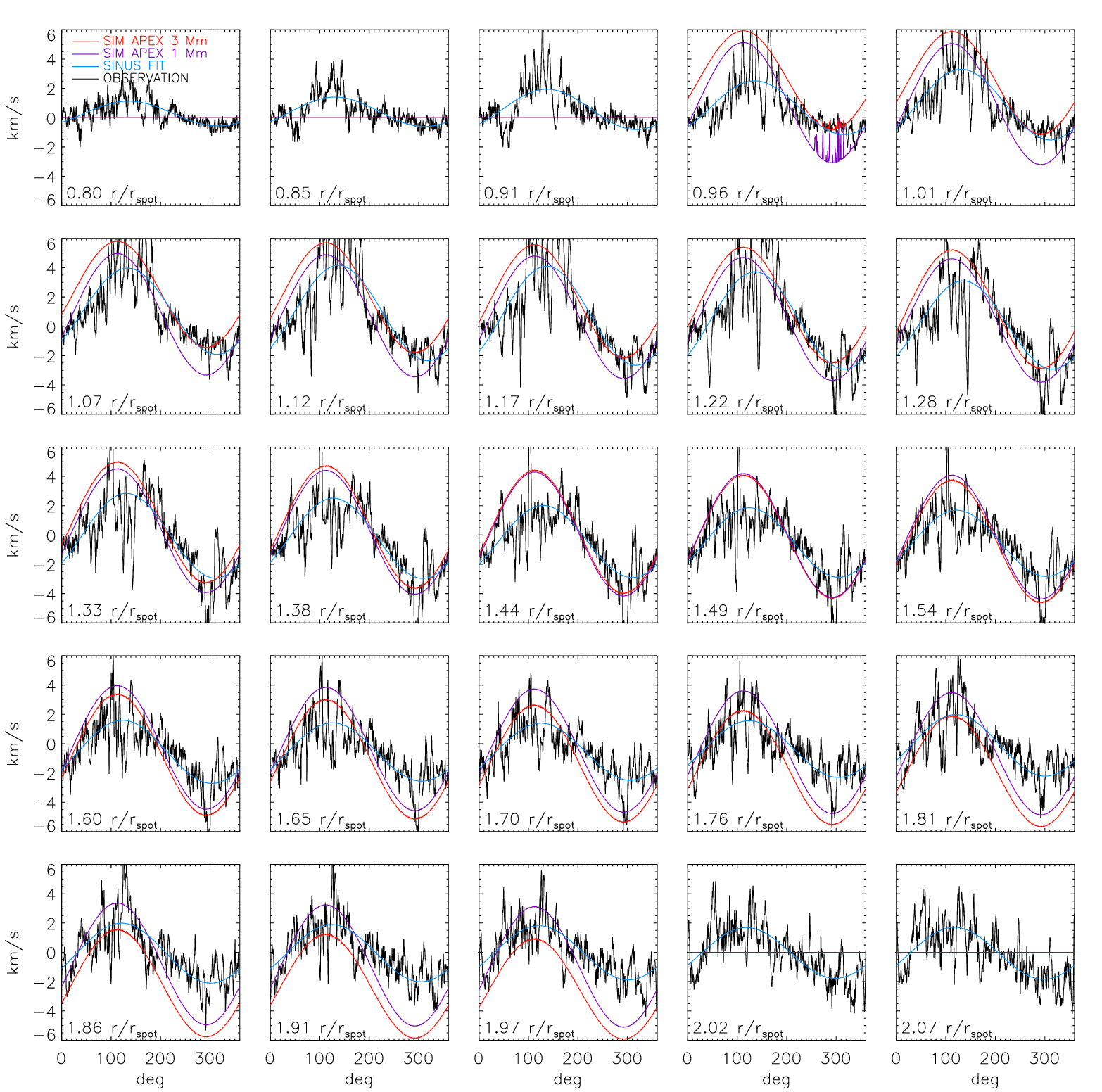}}$ $\\$ $\\
\end{minipage}
\caption{Fit of sinusoidal to LOS velocities on elliptical paths. Left: observed (left panel) and simulated velocity map (right panel) of sunspot No.~4 with overlayed elliptical paths. The black line points towards the limb. Right panels: observed velocity (black lines), fit of sinusoidal (blue lines), and simulated velocities for apex heights of 1\,Mm (purple lines) and 3\,Mm (red lines). The radial distance in fractional sunspot radii is given at the bottom of each sub-panel and increases from left to right and top to bottom.}\label{fig_fits1}
\end{figure*}

The resulting two-dimensional axisymmetric synthetic velocity maps $v^{LOS}_{(I)EF}(x,y)$ are then rotated around their center to match the location and image orientation of the observed sunspots on the solar disk, i.e., the maximal and minimal velocities have to be along the line connecting disc center and sunspot center in the observations. We derived the necessary rotation angle $\Psi$ from identifying the sunspot center of each observation in a full-disk HMI continuum intensity image and measuring its position angle relative to the equator. The rotated velocity maps $v^{LOS, ROT}_{(I)EF}(x,y)$ are then projected onto an ellipse through a coordinate transformation
\begin{eqnarray}
x^{PROJ} = x_{cen} + r \cdot ( a \cdot \cos \Phi \, \cos \Psi - b\cdot \sin \Phi \, \sin \Psi) \\
y^{PROJ} = y_{cen} + r \cdot (a \cdot \cos \Phi \, \sin \Psi - b\cdot \sin \Phi \, \cos \Psi) \,,
\end{eqnarray}
where $a=1$, $b = \cos \theta$, while $r$ is as defined above and the values of $x_{cen}$ and $y_{cen}$ compensate for the fact that the sunspot centers are not necessarily exactly in the middle of the IBIS FOV in the observations.

This provides synthetic velocity maps of the (I)EF LOS velocity for a sunspot at an arbitrary heliocentric angle $\theta$ and location $\Psi$ on the solar disk (see, e.g., Figure \ref{fig3}). The exact LOS velocity values are additionally a function of the selected flow speed $v_{(I)EF}(r)$, apex height $AP_{(I)EF}$ and for the IEF channels the location of the inner endpoint $r_0$. As can be seen in Figure \ref{fig3}, a variation of only the apex height leads to noticeable differences in the projected velocity maps, e.g., for $\theta = 39^\circ$ or 49$^\circ$ the areas of zero velocity (green regions) significantly change shape with the apex height, which offers the opportunity to use CLV observations of LOS velocities for diagnostic purposes. 

To match the varying true sunspot sizes in the different IBIS observations, the final projected synthetic velocity maps were just resized by a manually determined variable scaling factor. To verify that the projected velocity maps match to first order the actually observed sunspot shape, we always projected the continuum and H$\alpha$ line-core intensity of an observed round sunspot with about the same umbral and penumbral radius as the artificial sunspot in the (I)EF simulator in the same way as the synthetic velocities (see rightmost columns of Figure \ref{fig1} and the first two columns in Figure \ref{fig3}). 


\subsection{Sunspot Geometry}
The geometry of the sunspot in each observation was taken into account by determining three characteristic quantities: the sunspot center, the outer penumbral radius and the location of the symmetry line. The center of the sunspot was defined manually in each observation as the IBIS FOV was not necessarily centered on the sunspot, especially for the observations close to the limb. The sunspot radius $r_{spot}$ was defined as the radius of the circle in the simulated data that after the projection matched the outer penumbral edge (see, e.g., the red ellipses in Figure \ref{fig1}). The symmetry line of the sunspot indicates the direction from sun center through the sunspot center towards the closest limb position. It was derived from the location of the sunspot on HMI full-disk intensity images.

In addition, we visually determined the umbral, penumbral and superpenumbral radius for each sunspot in AIA 1700{\AA} images using the accumulations of network elements at the end of the moat cell for the latter (bottom three rows of Table \ref{tab_fitres}).
\subsection{Radial Cuts}
For each observation, we defined a set of 180 radial cuts with an angular separation of 2$^\circ$ in azimuth that started from the sunspot center and sampled the radial variation over 600 points with a spatial distance of 0\farcs063 each \citep[see, e.g.,][their Figure 1]{beck+choudhary2020}. On each cut, we determined the maximal velocity and the radial distance of its location along the cut. 
\subsection{Azimuthal Curves}
For the automatic derivation of flow speed and flow angle, we defined a set of 50 elliptical curves of increasing radius that were centered on the sunspot. The radial range was set to span 0.8--2.1\,$\times r_{spot}$ with a radial step width of 0.027\,$\times r_{spot}$ (left half of Figure \ref{fig_fits1}). The observed and simulated LOS velocities along those curves were extracted and evaluated under the assumption of an axisymmetric flow field using the Equations (2)-(5) of \citet{schlichenmaier+schmidt2000}. In that case, the LOS velocity $v(r,\phi)$ at a given radius $r$ and azimuth $\phi$ is given by
\begin{equation}
v(r,\phi) = a(r) \cdot \sin (\omega \phi + \psi(r) ) + b(r) \,, \label{eqsin}
\end{equation}
where $\omega = 1/2\,\pi$.

With $v_{vert}(r) = b(r)$ and $v_{hor}(r)  = a(r)$, the flow angle $\gamma(r)$ is given by
\begin{eqnarray} 
\gamma(r) = \arctan \left(\frac{v_{hor}(r)}{v_{vert}(r)\,\tan \theta}\right)\,,
\end{eqnarray}
and the absolute flow velocity $v_0(r)$ by
\begin{eqnarray} 
v_0(r) =  \frac{v_{vert}(r)}{\cos \theta \cos \gamma(r)}\,.
\end{eqnarray}

The velocity values along the ellipses were then evaluated by fitting the sinusoidal of Equation (\ref{eqsin}) to each curve (right half of Figure \ref{fig_fits1}). The IEF shows a lot of azimuthal fine-structure with multiple individual flow channels that do not necessarily all start at the same radial distance. The automatic fit was still able to match the observed velocity curves quite well, but only after the period $1/\omega$ was set to a fixed full period over 360$^\circ$.  

As a consistency check, we applied the same calculation to the projected LOS velocity maps from the simulator, but using for simplicity $v_{vert, SIM}(r) = <v_{SIM}(r,\phi)>_{|\phi}$ and $v_{hor, SIM}(r)  = max(v_{SIM}(r,\phi)-v_{vert, SIM}(r))_{|\phi}$ instead of a sinus fit.
\begin{figure*}
\resizebox{17.6cm}{!}{\hspace*{1cm}\includegraphics{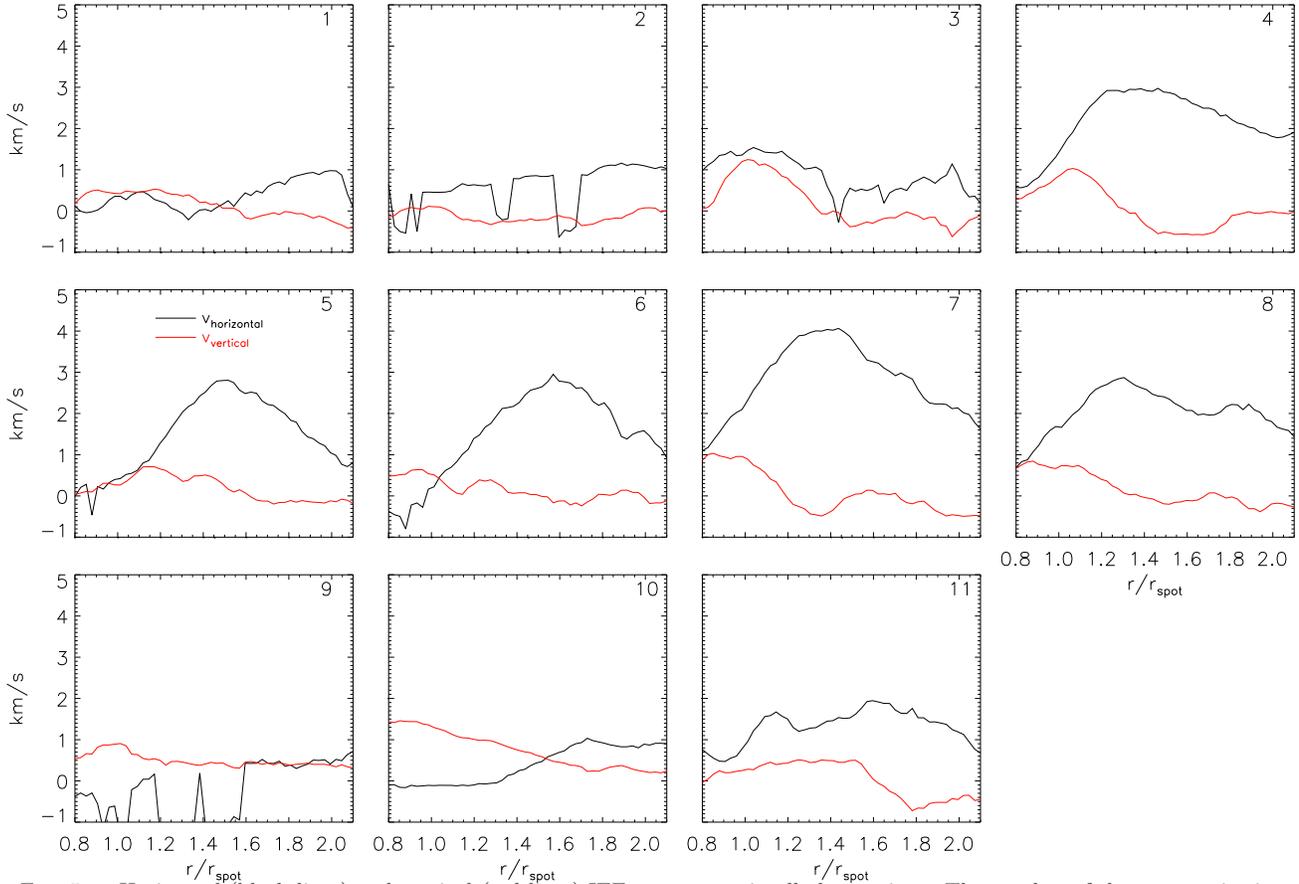}}$ $\\$ $\\
\caption{Horizontal (black lines) and vertical (red lines) IEF component in all observations. The number of the sunspot is given in the upper right corner of each panel.}\label{fig_horvertspeed}
\end{figure*}
\subsection{IEF Simulator ``Fit''}
The IEF simulator has three free parameters, the flow speed $v_{(I)EF}(r) = v_0 \equiv$ constant, the apex height of the loop $AP_{IEF}$, and the inner starting point of the IEF channels $r_0$. We initially tried to separately fit the simulator output to each sunspot observation with all three parameters being free. We used two different contributions to the $\chi^2$ to be minimized. The first was based on the squared difference between the observed and synthetic LOS velocity maps $v_{norm}(x,y) = v(x,y)/max(|v(x,y)|)$ normalized to their maximal values by
\begin{equation}
\chi^2_1 = \sum_{x,y} \left(v_{obs,norm}(x,y) - v_{SIM,norm}(x,y) \right)^2  \label{chi1}
\end{equation}
for all pixels $(x,y)$ where the simulator output was not zero. This contribution is independent of the exact flow speed in the observations or simulations, but measures the similarity in the spatial pattern.

The second contribution to $\chi^2$ was based on the value of the maximal velocity along all radial cuts $v_{max}(\phi) = max( |v(r,\phi)|)_{|r}$ by
\begin{equation}
\chi^2_2 = \sum_{\phi} \left(v_{obs,max}(\phi) - v_{SIM,max}(\phi) \right)^2 \,. \label{chi2}
\end{equation}
\begin{figure*}
\resizebox{17.6cm}{!}{\hspace*{1cm}\includegraphics{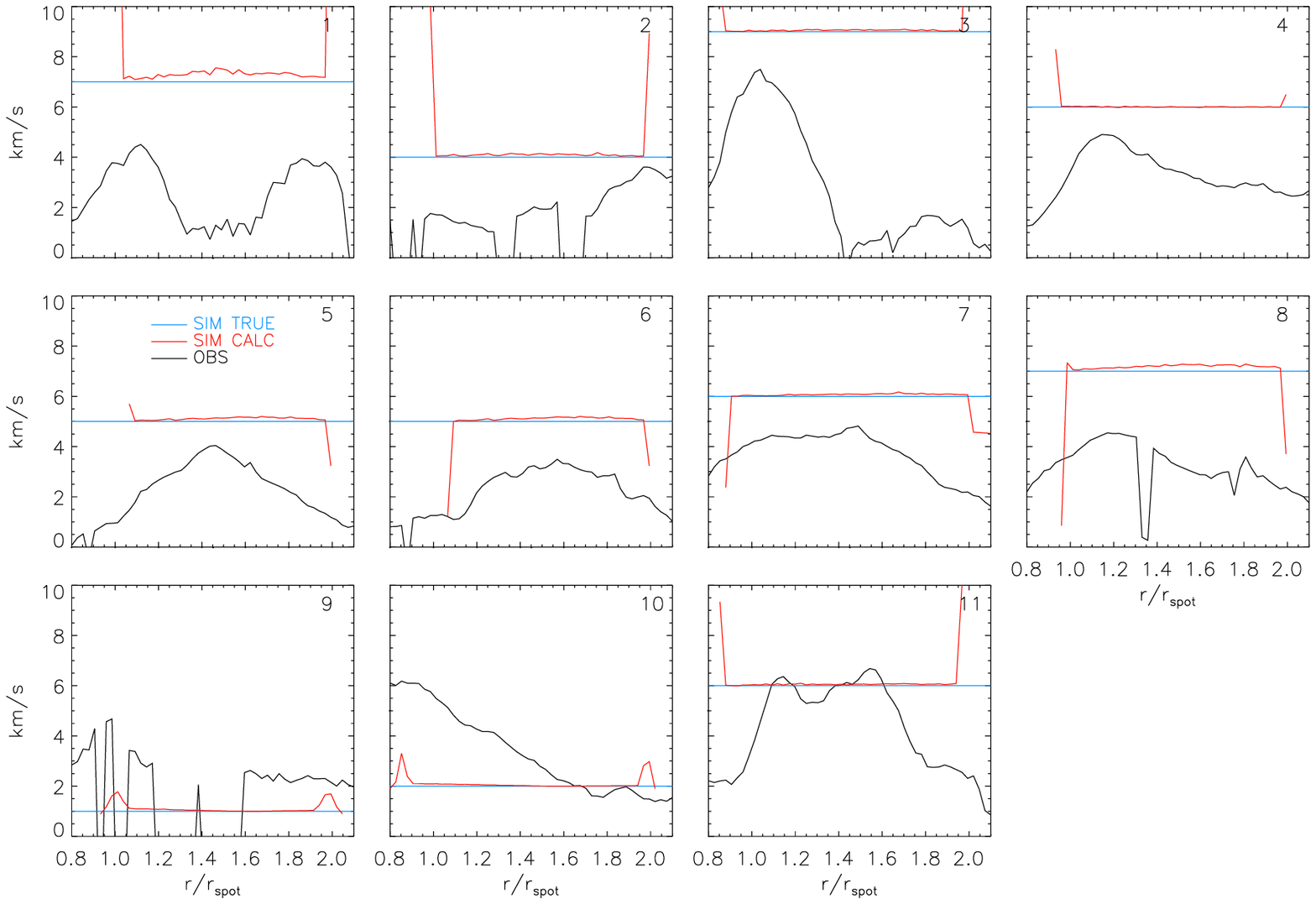}}$ $\\$ $\\
\caption{Absolute flow speed in all observations. Black lines: as derived from the observed LOS velocities. Red lines: as derived from the simulated velocities. Blue horizontal line: true input velocity in the simulator. The number of the sunspot is given in the upper right corner of each panel.}\label{fig_speed}
\end{figure*}

It turned out, however, that regardless of the combination of the two $\chi^2$ contributions, or when omitting the normalization by $v_{max}$ in $\chi^2_1$, the three free parameters of the simulator were ambiguous. The main limiting factor was the high degree of spatial fine-structure in the observed LOS velocity maps that show also flows that are unrelated to the IEF. We thus decided to first fix the value of $r_0$ for each sunspot because it is comparably well defined in the H$\alpha$ line-core velocity map of each observation (compare the 4$^{\rm th}$ and 8$^{\rm th}$ column of Figure \ref{fig1} or see Figure \ref{fig_bestfit} below). The corresponding $r_0$ values are listed in Table \ref{tab_fitres}. We also applied a spatial smoothing over a $5^{\prime\prime} \times 5^{\prime\prime}$ area to each observed velocity map to reduce the small-scale spatial variation.

We then ran the simulator for each sunspot over a sequence of apex heights $AP_{IEF}$ from 0.5 to 4.3\,Mm in steps of 0.2\,Mm and determined the best fit for the velocity $v_0$ based only on the $\chi^2$ contribution of Equation (\ref{chi2}) at each apex height. The velocity corresponding to the smallest of those 20 $\chi^2$ values was chosen as the final best-fit $v_0$ value for each sunspot, while the standard deviation of the best-fit velocities over the 20 runs was taken as the error estimate. Small values of the standard deviation below 0.5 km\,s$^{-1}$ (see Table \ref{tab_fitres}) indicate that a unique velocity value could be determined.

To determine the best-fit value for the apex height $AP_{IEF}$, we used the $\chi^2$ contribution as defined by Equation (\ref{chi1}) that is independent of the flow speed and is only based on the spatial patterns of relative velocities. We ran the simulator over the same discrete sequence of apex heights $AP_{IEF}$ as above and derived the best-fit value for $AP_{IEF}$ from an interpolation of the corresponding curve of $\chi^2$ values around its minimum.

Based on the fit results for $v_0$ and the maximal velocities in each observations, we then set the simulator speed $v_0$ to a fixed value for each sunspot (see Table \ref{tab_fitres}) for the generation of all plots in this paper that show simulator results.

\section{Results}\label{secres}
\subsection{Observed Inverse Evershed Flows}
The two lines of H$\alpha$ and \ion{Ca}{ii} IR show similar flow patterns in the superpenumbra (Figure \ref{fig1}). We initially planned to process both spectral lines, but given that the direct analysis only retrieves average properties and that the automatic comparison to the IEF simulator required a spatial smoothing, we decided to discard the \ion{Ca}{ii} IR measurements for the current study. All results in the following are therefore solely based on the line-core velocities of H$\alpha$. We plan to investigate multiple chromospheric spectral lines, adding \ion{He}{i} at 1083\,nm, and additional photospheric spectral lines for a subsample of the current sunspot selection in a subsequent study of the height dependence of the IEF.
\begin{figure}
\resizebox{8.8cm}{!}{\includegraphics{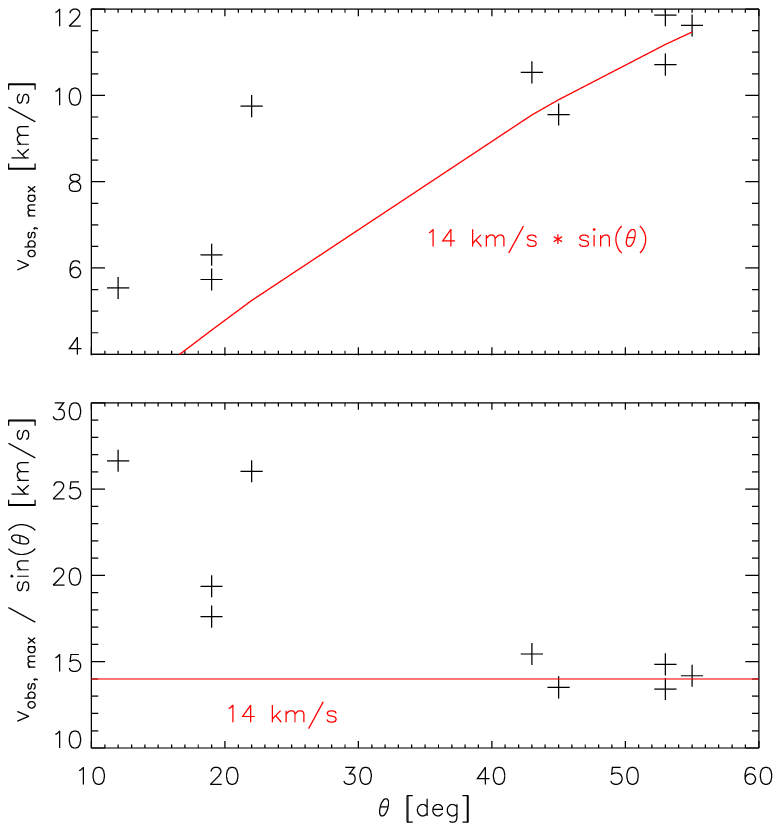}}$ $\\$ $\\
\caption{Maximal observed IEF velocities as a function of heliocentric angle $\theta$. Bottom panel: observed velocities divided by $\sin \theta$. Top panel: observed maximal velocities. The red lines indicate a velocity of $v_0=14$\,km\,s$^{-1}$ and $v_0 \sin \theta$, respectively.}\label{fig_helio}
\end{figure}
\subsubsection{Horizontal, Vertical and Absolute Flow Speed}
Figure \ref{fig_horvertspeed} shows the horizontal and vertical flow speed for all observations as derived from the sinusoidal fit. The vertical flow speed $v_{vert}$ shows downflows of about 1\,km\,s$^{-1}$ at the inner end for $0.8 < r/r_{spot} < 1.2$ that turn into upflows of about 0.5\,km\,s$^{-1}$ for $r/r_{spot}>1.2$. The horizontal flow component $v_{hor}$ usually increases from zero at $r/r_{spot}=1$ to about 2--3\,km\,s$^{-1}$ for $1< r/r_{spot}< 2$ with a maximum around $r/r_{spot}=1.2-1.6$. 
\begin{figure*}
\resizebox{17.cm}{!}{\hspace*{1cm}\includegraphics{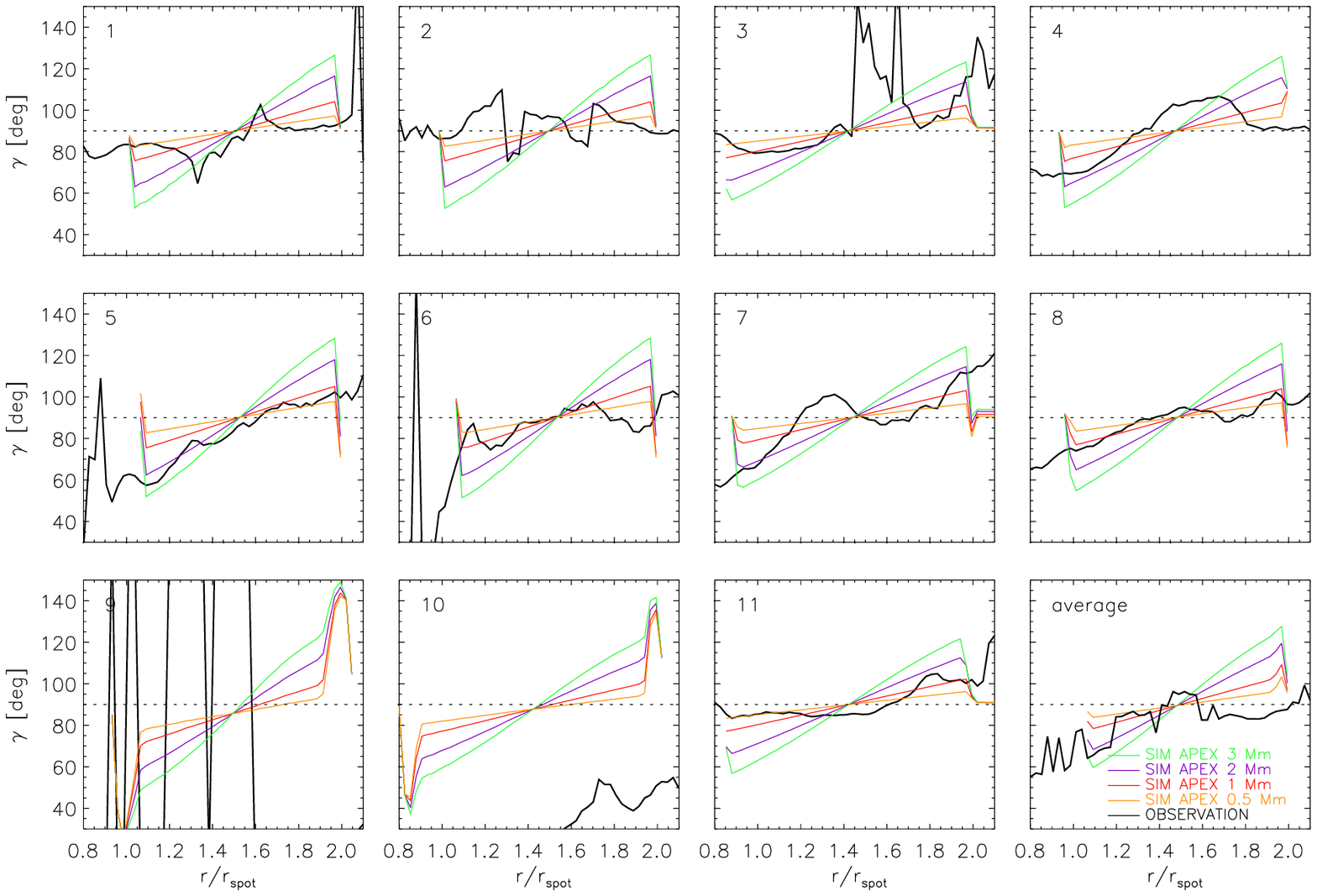}}$ $\\$ $\\
\caption{Inclination of the flow to the surface for all observations. Black lines: as derived from the observed LOS velocities. Orange, red, purple and green lines: as derived from the simulated velocities for an apex height of 0.5, 1, 2, and 3\,Mm. The number of the sunspot is given in the upper left corner of each panel. The panel in the lower right corner shows the average value of all observations.}\label{fig_inclination}
\end{figure*}

The absolute flow speed $v_0$ (Figure \ref{fig_speed}) is somewhat smoother in the radial direction with values around 4\,km\,s$^{-1}$ for $1<r/r_{spot} < 1.8$. All values correspond to azimuthal averages and are below the chromospheric sound speed. The values of $v_0$ selected for the simulator are generally larger than the result from the sinus fit to the observations, but the observed velocities on the elliptical curves show a large fluctuation with common occurrences of flow speeds above the amplitude of the fitted sinus curve (right-hand side of Figure \ref{fig_fits1}). The small differences between the red and blue lines in Figure \ref{fig_speed} stem from the discrete spatial and azimuthal sampling of the simulated velocities on the elliptical curves, while the jumps at the start and end of the red lines result from missing data points in the projected simulator image on the innermost and outermost ellipses.

As the fit of the sinusoidal does not necessarily recover the maximal velocity amplitudes because of the large spatial variation, we also determined only the maximal absolute LOS velocity in each observation in the area where the projected simulator velocity map exhibits non-zero values, regardless of azimuth and radial distance. This generally isolates the IEF from flows of a different origin. Those numbers provide a clearer picture of the variation of the flow speed with heliocentric angle $\theta$ (Figure \ref{fig_helio}). This trend with $\theta$ can also be already seen in the velocity maps of Figure \ref{fig1} that are all displayed within the same velocity range, e.g., by comparing observation No.~1 or 2 with No.~7 or 8. The maximal velocity increases with the heliocentric angle (Figure \ref{fig_helio}) and matches to first order to a curve of $v_{max}(\theta) = v_0 \cdot \sin \theta$ with $v_0 = 14$\,km\,s$^{-1}$. This type of dependence on $\theta$ would result for a purely horizontal flow of constant and identical velocity for all sunspots. The fact that the observed maximal flow speeds for $\theta < 30^\circ$ are larger than the simple curve complies with the fact that the IEF deviates from horizontal near its downflow points \citep{choudhary+beck2018}. 

\subsubsection{Average Flow Angle}
Figure \ref{fig_inclination} shows the flow angle for all observations together with the corresponding value in the simulator for apex heights from 0.5 to 3\,Mm. As expected from the vertical velocities of the previous section, the inclination to the local surface normal changes from below 90$^\circ$ for $r/r_{spot} < 1.4$ to above for larger radial distances. The minimum inclination is about 60$^\circ$, while reliably determined maximal values are about 110$^\circ$. An apex height of 2\,Mm gives the best match to the observed radial variation for most sunspots, while apex heights of 0.5 and 3\,Mm are too shallow or too steep, respectively. Table \ref{tab_inc} lists the inclination values in the simulator for an apex height of 2\,Mm. 
\begin{figure*}
\resizebox{17.cm}{!}{\hspace*{1cm}\includegraphics{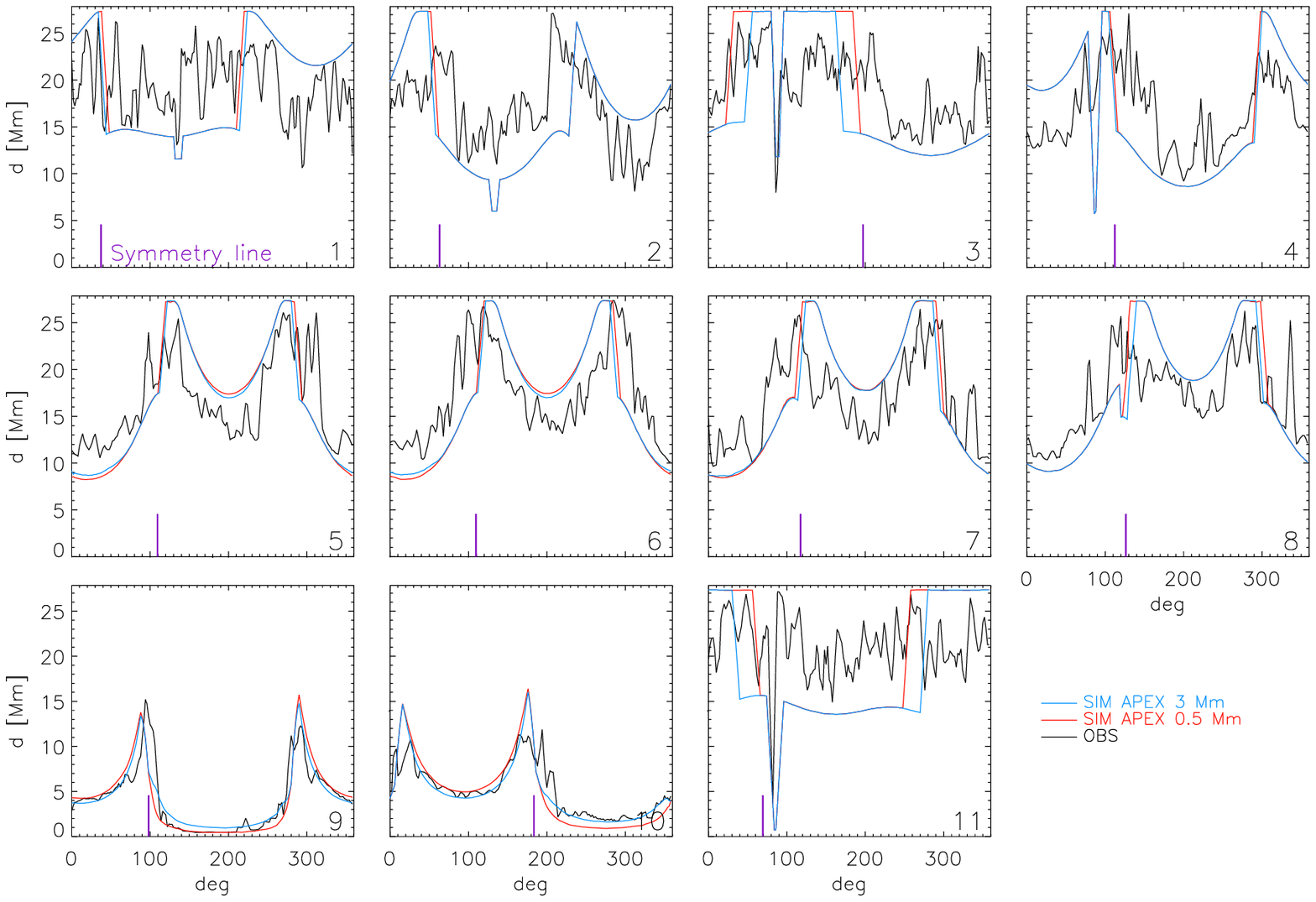}}$ $\\$ $\\
\caption{Radial distance of maximal unsigned velocities. Black line: from the observed LOS velocities. Blue/red lines: from simulations with an apex height of 3\,Mm and 0.5\,Mm, respectively. The short vertical purple bar indicates the location of the symmetry line. The number of the sunspot is given in the lower right corner of each panel}\label{fig_posmax}
\end{figure*}

\subsubsection{Location of Downflow Points}
The radial distance of the locations with maximal velocity on the radial cuts (Figure \ref{fig_posmax}) can be used to verify the choice of the inner start point $r_0$ in the simulator and the general match of the geometry. The match between the positions of maximal velocity in the observations and the simulator is close, with generally a smaller distance on the center than on the limb side as expected. Somewhat surprisingly, the biggest variation of the position is seen near the symmetry line where the LOS velocity approaches zero. A visual inspection of the synthetic maps in Figure \ref{fig3} ensures that the effect is real and to be expected because the flow velocities recede rapidly from the sunspot when approaching the symmetry line in azimuth. It is still noteworthy that this trend around the symmetry line is well matched for all observations, and especially also for the two sunspots Nos.~9 and 10 at heliocentric angles above 70$^\circ$ where, e.g., the sinus fit failed to converge. Unfortunately, the radial distance is only weakly sensitive to the apex height, as the two curves for 0.5 and 3\,Mm are nearly identical at all heliocentric angles. 
\subsubsection{Loop Geometry}
\begin{figure}
\resizebox{8.8cm}{!}{\hspace*{1.5cm}\includegraphics{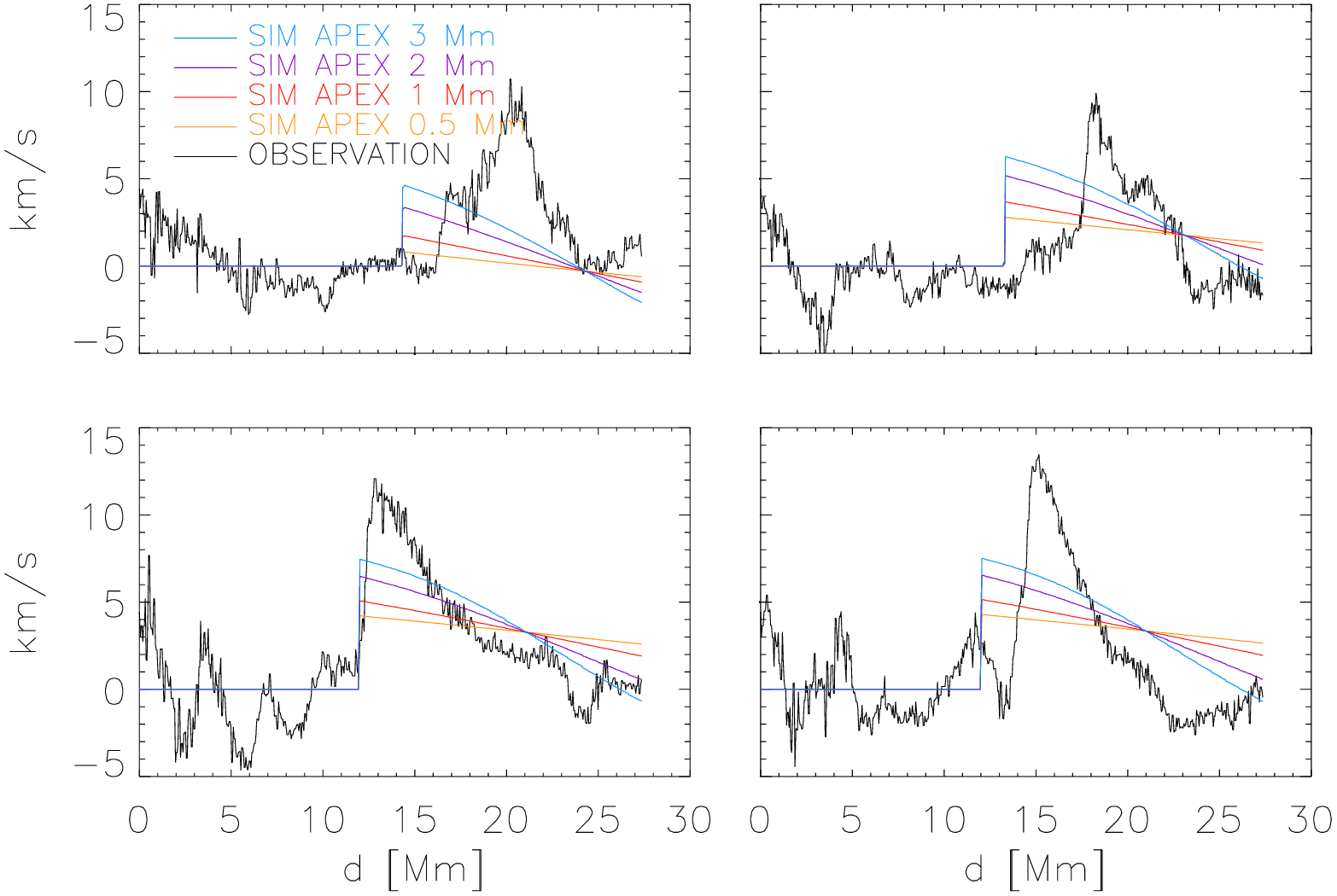}}$ $\\$ $\\
\caption{Velocities along four radial cuts in sunspot No.~3. Black lines: observed velocities. Orange, red, purple, blue lines: simulation with apex heights of 0.5, 1, 2, and 3\,Mm. The observed velocities have a downflow point at an arbitrary radial distance.}\label{fig_radial_cuts}
\end{figure}

Even if the simulator output shows an acceptable match to the observations on average (Figures \ref{fig_inclination} and \ref{fig_posmax}), the assumed parabolic shape of the loops in the simulator is presumably at odds with reality. Figure \ref{fig_radial_cuts} compares the observed and simulated velocities on four radial cuts in observation No.~3. The observed velocities drop much faster over a shorter distance from the downflow point than the simulator predicts, even for the case of the steepest loops with 3\,Mm apex height. We verified that this is the same for all observations. The two possible explanations would be that either the flow speed is not constant but accelerates toward the downflow point, or that the true loops are steeper than in the simulator. The best-fit values for the apex height limit it to 3\,Mm at maximum and the total flow speed (Figure \ref{fig_speed}) does not show strong acceleration, which makes the latter explanation more likely.
\begin{table*}
\caption{Manually set, fitted or directly derived characteristics of each observation}\label{tab_fitres}
\begin{tabular}{c|ccccccccccc|c}\hline\hline
Spot No. &  1 & 2 & 3 & 4 & 5 & 6 & 7 & 8 & 9 & 10 & 11 & units\cr\hline
$\theta$   &  12  & 19 & 22 & 45 &  53 & 53  &  55 & 43 & 79 & 76  & 19 & degree\cr\hline
& \multicolumn{11}{c}{Manually set} & \cr\hline
$r_{spot}$ & 15.1 & 14.0 & 19.7& 16.4& 17.8 & 17.8 &  20.3 & 18.7 & 10.1& 10.1 & 19.7 & Mm  \cr
$r_0$   &  1.02  & 1.00 & 0.85 & 0.95  & 1.07 &  1.08 & 0.90 & 0.98 & 1.00 & 0.85 & 0.85 & $r/r_{SPOT}$\cr
$|v_{SIMU}|$ &  7 &  4 &  9 & 6 & 5 &  5 & 6 & 7 & 1 & 2 & 6  & km\,s$^{-1}$\cr\hline
 &   \multicolumn{11}{c}{Fit of simulator to maximal velocities} & \cr\hline
$|v_{FIT}|$  & 6.0 &  3.0 & 10.5 & 4.7 &  4.9  & 5.5 &   6.7&  5.4 & --$^1$ & 4.3 & 7.9  & km\,s$^{-1}$\cr
rms $|v_{FIT}|$  & 1.3 &  0.4 & 1.7 &  0.5 & 0.1 & 0.2 & 0.2 & 0.5 &  --$^1$ & 0.1 & 2.1 & km\,s$^{-1}$ \cr\hline
 &   \multicolumn{11}{c}{Fit of simulator to spatial map} & \cr\hline
apex height   & 0.96 &  1.19  & 1.81 & 2.38 & 3.23 &  3.25 & 3.80 & 2.45 & --$^1$ & 2.00 & 1.59  & Mm \cr\hline
& \multicolumn{11}{c}{Directly derived from observations} & \cr\hline
$|v_{MAX}|$ & 5.5 & 5.7 & 9.8 & 9.6 & 10.7 & 11.9 & 11.6 & 10.5 & 4.0 & 6.4 & 6.3 & km\,s$^{-1}$ \cr
$<B>$ & 2431 & 2514&2808 & 2823 & 2714 & 2716 & 2623 &2531 & -- & -- & 2670 & G \cr
$B_{max}$ & 2693 & 2638 & 3272 & 3399 & 3077 & 3148& 3171 & 3053 & -- & -- & 3515 & G\cr
$r_{UMBRA}$ & 7.2 &  6.4  & 9.2  &  8.0 & 8.2 &   8.2 & 10.6 &  8.0 &   5.7 & 5.0 &  9.6 & Mm\cr
$r_{PENUMBRA}$ & 13.8 &  13.5 & 17.0 &  14.8 &  14.5 &  14.5 & 17.8 &  16.0 &  11.8 & 8.1&  20.3 & Mm\cr
$r_{SUPERPU}$ & 22.6& 22.5& 26.5& 23.5& 24.3&  24.3& 27.1& 27.7&20.5& 13.5& 27.2 & Mm\cr
\end{tabular}\\
$^1$: no convergence
\end{table*}

\begin{table*}
\caption{Flow angle as a function of radial distance for an apex height of 2\,Mm}\label{tab_inc}
\begin{tabular}{c|cccccccccccc}\hline\hline
$r/r_{SPOT}$ & 1.00 & 1.07 & 1.15 &  1.22 & 1.30 &  1.37 &1.45 & 1.52 & 1.60 &  1.67 &  1.75 &  1.82 \cr
$\gamma$ &  71 &  73 & 75 &79 &84 &88 &93 &98 & 102 &106 & 110 &116 \cr
\end{tabular}
\end{table*}

\subsection{IEF Simulator}
\begin{figure*}
\resizebox{14.cm}{!}{\hspace*{1cm}\includegraphics{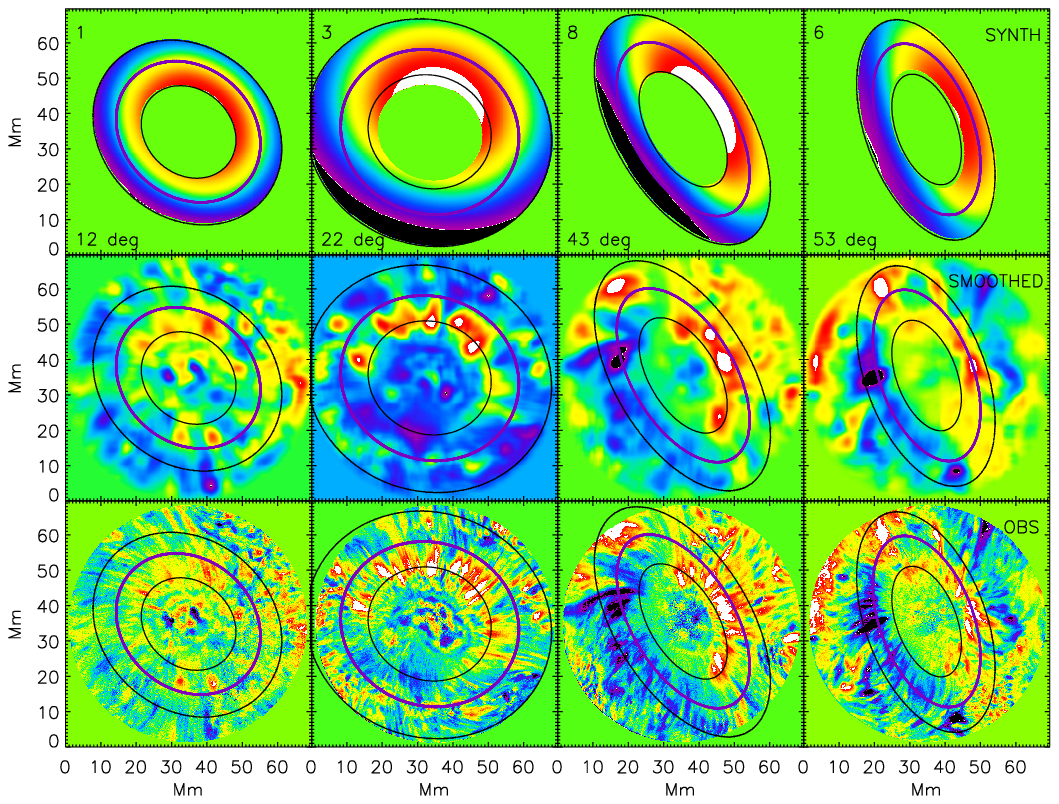}}\hspace*{.5cm}\includegraphics{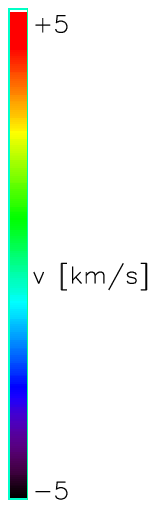}$ $\\$ $\\$ $\\
\caption{Examples of the best-fit solution. Bottom row: observed velocity maps of (left to right) observations No.~1, 3, 8 and 6 sorted by increasing heliocentric angle. The orientation of each observation was modified to have the center side always towards the upper right corner. Second row: the same after spatial degradation. Top row: corresponding simulator output using a fixed apex height of 2\,Mm and optimized flow speeds. The black ellipses indicate the inner and outer penumbral boundary in the simulator. The purple ellipses indicate where the inclination in the simulator passes through 90$^\circ$.}\label{fig_bestfit}
\end{figure*}
\begin{figure*}
\resizebox{16cm}{!}{\includegraphics{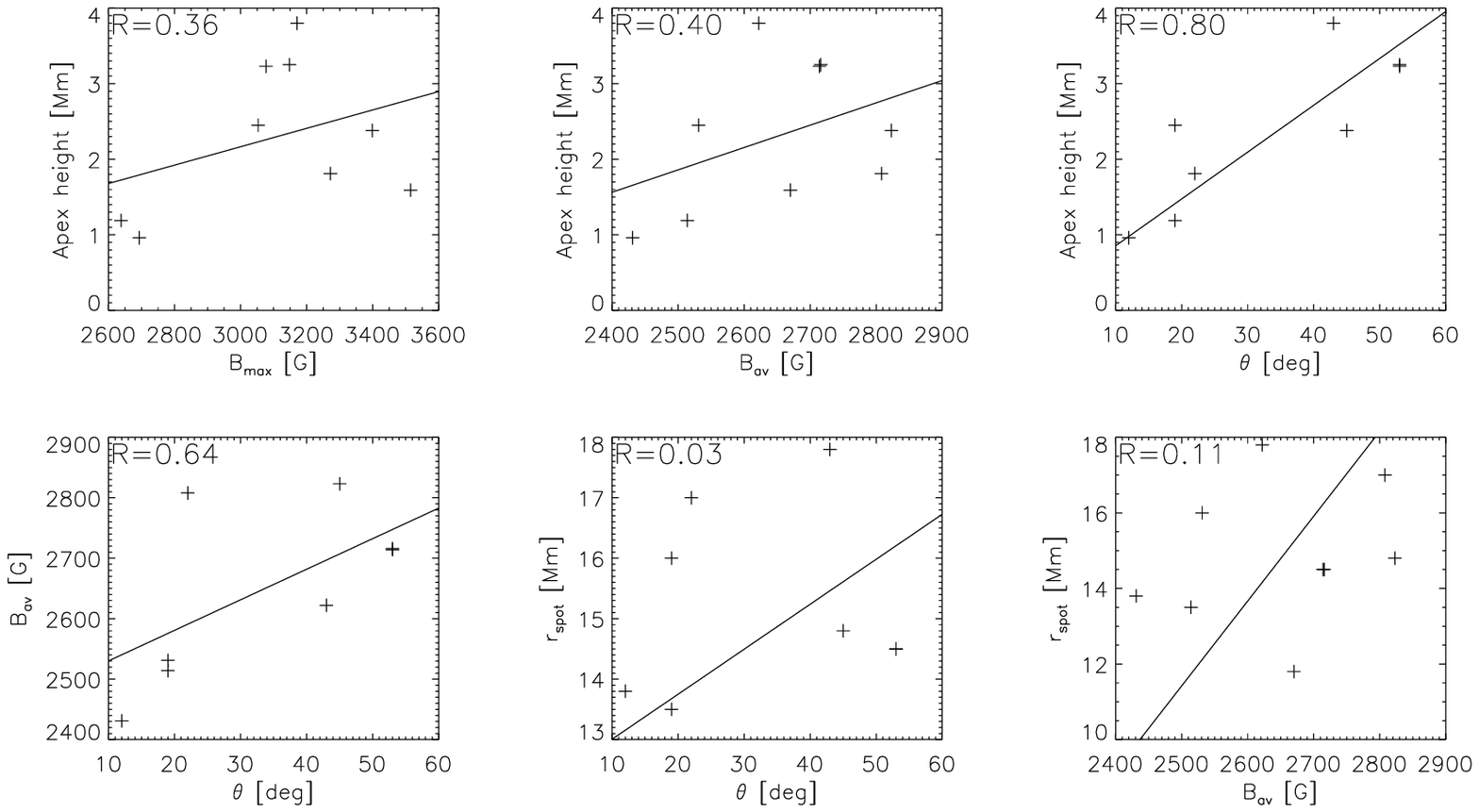}}
\caption{Relations between apex height, umbral field strength, sunspot radius and heliocentric angle. Bottom row, left to right: average umbral field strength $B_{av}$ vs.~heliocentric angle $\theta$, sunspot radius $r_{spot}$ vs.~$\theta$ and $r_{spot}$ vs.~$B_{av}$. Top row, left to right: apex height $AP$ vs.~maximal umbral field strength, $AP$ vs.~$B_{av}$ and $AP$ vs.~$\theta$. The inclined straight lines indicate the regression line to the data points, while the value of the linear correlation coefficient $R$ is given in the upper left corner in each panel.}\label{fig_bvsapex}
\end{figure*}
\subsubsection{General}
The IEF simulator fulfills its purpose to reproduce average properties of observed LOS velocities at different heliocentric angles. The high spatial resolution of the observations actually hinders the match. Figure \ref{fig_bestfit} shows four observations at increasing heliocentric angle together with the spatially smoothed maps that were used in the fit of simulator parameters. The axisymmetric IEF simulator can match the observations only when reducing the spatial resolution and smearing out the azimuthal variation at the price of a reduced flow velocity. 

In the direct comparison of the IEF simulator output and the observed velocity maps, it seems that especially for large sunspots the assumed length of the simulated IEF fibrils is too short (third and fourth column of Figure \ref{fig_bestfit}) because the observed IEF pattern extends beyond those in the projected simulator images. The IEF simulator was set to use ratios of 1:2:4 for the umbral, penumbral and superpenumbral radius. The measured values (Table \ref{tab_fitres}) give corresponding ratios of 1:1.89:3.05 on average, which implies that the choice for the simulator is justified in general, but might be off for individual sunspots. 

The IEF simulator answers better than the observations the question whether upflows and downflows relative to the solar surface show up as blue and red shifts, respectively. This identification of the true flow direction relative to the surface from the sign of the LOS velocity only holds for sunspots at a heliocentric angle $<30^\circ$, e.g., the ring of downflows at the inner penumbral boundary for the left two columns in Figure \ref{fig_bestfit}. The middle purple ellipse in each panel indicates where the flow angle in the simulator passes 90$^\circ$ relative to the surface normal. In the simulator velocities and most of the observations the sign of the LOS velocity does not change at that place for sunspots with $\theta > 30^\circ$.      
\subsubsection{Best-fit Solution}\label{sim_results}
The automatic fit of the IEF simulator to the observations was only partly possible. We encountered a few unresolvable ambiguities due to the large spatial variation of the observed velocities that forced us to fix some parameters such as the inner radial footpoint or to remove the actual velocity value from the fit keeping only the relative spatial distribution. 

Figure \ref{fig_posmax} shows that the geometry of the flow field is in general well recovered. Even for the sunspots close to the limb the locations of maximal velocities are well matched, where the symmetry line with nearly zero velocities turned out to be a major diagnostics. 

The main free parameter in the fit is the apex height that varies from 1 to 3\,Mm (Table \ref{tab_fitres}). The analysis of observations of the same sunspot yielded a similar value of the apex height each time (observations Nos. 1 and 2, and 5 and 6). The inclination values to the surface normal in Figure \ref{fig_inclination} suggest that a value of 2\,Mm gives the best fit across all sunspots. Figure \ref{fig_radial_cuts} demonstrates that the IEF fibrils should be steeper near the footpoints and presumably flatter at the apex than the assumed parabolic shape. The observed flow speeds drop much faster over a shorter distance than the simulator even at an apex height of 3\,Mm. The flow speeds retrieved by the simulator fit only can match average values and usually underestimate the maximal observed velocities by a factor of two.

\subsection{Relations between Field Strength, Sunspot Size, Apex Height and Heliocentric Angle}
Our data sample is not uniform, but consists of different sunspots observed at different stages of their evolution and at different heliocentric angles $\theta$. Table \ref{tab_fitres} lists the average and maximal umbral field strength of each sunspot. To verify the impression from the table that there is a relation between the field strength and the apex height, we plotted the umbral field strength values against the apex height (top row in Figure \ref{fig_bvsapex}). Even if there seems to be a dependence of increasing apex height with increasing umbral field strength, it is difficult to assess its significance. There is both a positive correlation between apex height and heliocentric angle (upper right panel of Figure \ref{fig_bvsapex}) and heliocentric angle and umbral field strength (bottom left panel of Figure \ref{fig_bvsapex}). As expected, there is no correlation between sunspot size and heliocentric angle (bottom middle panel of Figure \ref{fig_bvsapex}), while the sunspot size shows a weak dependence on the umbral field strength (bottom right panel of Figure \ref{fig_bvsapex}).

While the apex height is derived from diagnostics in chromospheric lines that can well show a dependence on projection effects with increasing $\theta$, the umbral field strength was derived from the photospheric \ion{Si}{i} line and thus should not depend too strongly on $\theta$. If the formation height of the \ion{Si}{i} line would shift upwards with increasing $\theta$, it should show the opposite trend of lower field strengths at larger heliocentric angles.  With our current sample of sunspots we can thus not clearly distinguish between projection effects and a dependence of IEF properties on umbral field strength because the lower left panel of Figure \ref{fig_bvsapex} suggests a bias in the sample. 

\section{Discussion}\label{secdisc}
The IEF occurs along chromospheric fibrils that mostly connect the outer penumbra with the superpenumbral or moat boundary. Knowing the exact properties of these flows and the corresponding flow channels would help in understanding the global sunspot structure. In other words, it could reveal the relationship of the surrounding moat region with the main umbral-penumbral structure. The length, height, inclination, flow speed, and the location of the apex with respect to the penumbral boundary are critical parameters to pinpoint the driving mechanism that sustains the quasi-steady IEF with life times from 10 to more than 60 min \citep{maltby1975,dialetis+etal1985,beck+choudhary2020}. Our high-resolution observations show that individual flow channels are highly structured and flows in and between them vary widely. Our comparison of spatially smoothed velocity observations with a simplified model only gives average parameters. 
\subsection{Flow Angle and Flow Speed}
The inclination of the flow channels varies among different sunspots and among individual fibrils within a sunspot, mainly as a function of the radial distance. Our previous study showed an inclination of -27$^\circ$ to the local horizontal at the inner end of IEF channels \citep{beck+choudhary2019}, while the current study predicts a smooth variation from -20$^\circ$ to +26$^\circ$ in the radial direction with downflows at the inner and upflows at the outer end. The flow speeds are about 4 to 6 km\,s$^{-1}$ on average with resolved peak values of up to 12 km\,s$^{-1}$. \citet{georgakilas+etal2003} found inclinations of -30$^\circ$ to +6$^\circ$ with velocities of about 6 km\,s$^{-1}$, while \citet{haugen1969} found inclinations of -55$^\circ$ to +70$^\circ$ with velocities of 7 km\,s$^{-1}$. All studies that derived the radial variation of the flow angle under the assumption of axial symmetry of the sunspot find downflows at the inner and upflows at the outer end. Our CLV study revealed that these cannot be identified just from the sign of the LOS velocity as blue or red shifts for sunspots at a heliocentric angle larger than about 30$^\circ$.

The average IEF flow speeds are apparently subsonic. Spatially fully resolved IEF channels reach the chromospheric sound speed, while all velocity determinations were shown to underestimate the true velocities by a factor of about two in addition because the flow channel only causes a weak line satellite in the wing \citep{choudhary+beck2018,beck+choudhary2019,beck+choudhary2020}. We find a less pronounced radial variation of the absolute flow speed than in \citet{alissandrakis+etal1988} or \citet{dere+etal1990}, but it is not possible to exclude any possible acceleration towards the inner end with the current average results.

There is, however, a wide variation of flow speeds in individual channels, which are sometimes much higher than the average speed. The shock formation at the termination of the IEF channels near the outer edge of the penumbra could be due to these observed higher speeds or because of encountering the reduced local sound speed as the plasma enters the lower atmosphere with its high density. Another mechanism could be the funneling effect, i.e., more expanded chromospheric magnetic field structures contract in diameter at lower heights \citep{spruit1976,montesinos+thomas1989}, which could increase the flow speed near its termination to beyond sonic speeds.
\subsection{IEF Loop Geometry}
The apex position and apex height of the assumed loops are two other parameters that characterize the flow channels. Our modeling of the flow pattern for the CLV observations yielded that the H$\alpha$ flow channels connecting penumbra and superpenumbra seem to have a somewhat variable apex height between 1 and 4 Mm, while on a global average a height of about 2\,Mm gives the best match. The simplified modeling did not provide independent information on the radial distance of the apex, since the fibril length of 10 to 20\,Mm was fixed as proportional to the sunspot's penumbral radius with the apex being reached exactly in the middle. Previous studies estimated the apex height at the upper end of our range with 4--5\,Mm \citep{maltby1975,dialetis+etal1985}. A comparably low apex height or an extended vertical extension of the IEF channels to lower heights would be required by the obvious presence of the IEF channels up to the end of the moat in the \ion{Ca}{ii} IR line (Figure \ref{fig1}) that samples lower layers than H$\alpha$. 

As can be seen in Figure \ref{fig_bvsapex}, the apex height could depend on the umbral magnetic field strength of the sunspot and the heliocentric angle. \citet{dialetis+etal1985} noted an increase in flow speed with height within the H$\alpha$ line. For our CLV study that effect could couple into the rise of the formation height of especially chromospheric spectral lines for increasingly inclined LOSs. With our current sunspot sample, we cannot disentangle the relative contributions of both effects. The optimum sample would cover the same sunspot at different heliocentric angles, with best a constant umbral field strength during the disc passage. Similar to \citet{maltby1975}, we find that the assumed parabolic shape might be slightly off. The rapid variation of the flow speed near the downflow points indicates that a curve with a flatter top and steeper ends might be more realistic. 
\subsection{Variation of IEF Properties between Sunspots \& Model Limitations}
We find in general a large variation of the IEF properties for a single sunspot and between different sunspots. While sunspots at the same heliocentric angle show similar maximal, horizontal and total flow speeds (Figures \ref{fig_horvertspeed} and \ref{fig_speed}), individual IEF channels show a comparably large fluctuation in their properties. The wide range of inclinations and flow speeds among and within the flow channels is likely dictated by the corresponding local magnetic field lines and their specific connectivity. Apart from the observations close to disc center, all sunspots comply with a maximal true flow speed of 14\,km\,s$^{-1}$, while the former exceed the predicted speed after projection to the LOS. The sunspots at intermediate heliocentric angles 40--60$^\circ$ are significantly larger than the rest of our sample. They all show a much clearer spatial signature of IEF flows with in general longer IEF fibrils.

In our IEF simulator, most of the geometry is predefined by the constant ratios of umbra, penumbra and superpenumbra. \citet{sobotka+roudier2007} found a dependence of those radii on the sunspot age and evolution stage instead. As mentioned above, the shape of the IEF channels might not be well described by a parabola. While the IEF simulator can to first order reproduce average properties across different sunspots with a sort of common set of parameters, it seems that a better match requires an explicit adjustment to individual sunspots, and even further to individual flow channels in a single sunspot. This could be easily included in the current model by replacing the assumption of axisymmetry with a flexible radial flow speed profile, loop geometry and starting or end point, which could to some extent be done automatically based on a previous analysis of the observed flow pattern with azimuthal resolution. One major ingredient that is currently missing is the actual loop geometry and connectivity. We plan to use magnetic field extrapolations \citep[e.g.,][]{yadav+etal2019} for a few of the sunspots in our sample to obtain a better matching input geometry for the IEF simulator.

\section{Conclusions}\label{secconcl}
A simplified model of arched loops with a parabolic shape, 10--20\,Mm length, an apex height of 2--3\,Mm and a true constant flow speed of 10--14\,km\,s$^{-1}$ can on average reproduce the flow field of the inverse Evershed flow in the chromosphere in the superpenumbra at different heliocentric angles. Each sunspot shows a high degree of spatial variation in flow properties in speed, location, apex height and length of the flow channels. Those properties seem to depend strongly on general characteristics of the sunspot such as the average umbral field strength or the sunspot size, and on local characteristics like the shape and connectivity of individual magnetic field lines. The sign of the LOS velocity can only be used to distinguish between upflows and downflows  relative to the solar surface in IEF channels for sunspots at a heliocentric angle of less than 30$^\circ$.

A more consistent data sample with observations of sunspots of similar size or of the same sunspot at different heliocentric angles would be needed to fully separate projection effects from the true variation of the flow geometry. Observations with a low spatial resolution of 1$^{\prime\prime}$ could be used for this purpose at the price of underestimating the true flow speeds. A model of a siphon flow of constant sonic speed along elevated loops with a flattened top would match the observed characteristics of the IEF in the chromosphere at different heliocentric angles.      

\begin{acknowledgements}
The Dunn Solar Telescope at Sacramento Peak/NM was operated by the National Solar Observatory (NSO). NSO is operated by the Association of Universities for Research in Astronomy (AURA), Inc.~under cooperative agreement with the National Science Foundation (NSF). HMI data are courtesy of NASA/SDO and the HMI science team. IBIS has been designed and constructed by the INAF/Osservatorio Astrofisico di Arcetri with contributions from the Universit{\`a} di Firenze, the Universit{\`a}di Roma Tor Vergata, and upgraded with further contributions from NSO and Queens University Belfast. This work was supported through NSF grant AGS-1413686.
\end{acknowledgements}
\bibliographystyle{aa}
\bibliography{references_luis_mod}

\end{document}